# Defining the optimal level of business benefits within IS/IT projects: Insights from benefit identification practices adopted in an IT Service Management (ITSM) project

**Ravinda Wijesinghe**
Department of Information Systems and Logistics
Swinburne Business School
Swinburne University of Technology
Hawthorne, Australia
Email: rwijesinghe@swin.edu.au

**Helana Scheepers**
Department of Information Systems and Logistics
Swinburne Business School
Swinburne University of Technology
Hawthorne, Australia
Email: hscheepers@swin.edu.au

**Stuart  Mcloughlin**
Department of Information Systems and Logistics
Swinburne Business School
Swinburne University of Technology
Hawthorne, Australia
Email: smcloughlin@swin.edu.au

## ABSTRACT

*The popularity of benefit realization management (BRM) in today's IT-enabled world is fast gaining traction within IT organisations around the world. However, there appears to be limited attention paid to the intra-organisational practice by which benefits are identified. The purpose of this paper is twofold: firstly, to describe and define a practice approach that reflects the ongoing organisational investment through a benefit identification process that serves to exploit a number of benefit identification methods. Secondly, to underline and reflect on the importance of this benefit identification process in the context of IT service management (ITSM). This is achieved through a case study of an information technology infrastructure library (ITIL) implementation in a multi-national organization. The case study exposes a pragmatic practice approach of customising such implementations in an effort to achieve a cost effective implementation of IT services that reflects the context and requirements of that specific organisation.*

## 1. INTRODUCTION

The realization of benefits through IS/IT investments that results in organizational change and development is one of the most frequent and critical business and policy issues being discussed and researched in the domain of information systems (Ward & Daniel, 2012). The continued failure of high profile IT projects around the world has put significant challenges on organizational executive teams to properly execute their IS/IT driven change projects so as to realize value through the achievement of the projected optimum level of business benefits (Gingnell, Franke, Lagerström, Ericsson, & Lilliesköld, 2014). The popularity of benefits-driven approaches in IS/IT investments has significantly risen in recent years in concert with other popular mechanisms such as the balanced scorecard (Gacenga, Cater-Steel, Toleman, & Tan, 2011). This is both a function and consequence of the need to continuously involve impacted stakeholders and the necessary ongoing re-evaluation of business objectives with IS/IT changes.

Benefit driven approaches are based on the premise that, given IT has no inherent value, benefits will only arise as and when IT enables people to do things differently (Ward & Daniel, 2012). For this reason, the starting point of any benefit-driven approach is to consider the impacted stakeholders, what they are going to do differently in the context of new investment and how this is of benefit to them. This focus on benefit identification is particularly important because potential changes may impact significantly on the relationships, roles and working practices of many stakeholders across the organization. As benefits come from organizational change that enables people to do things differently, a key implication is that there is a need to ensure business ownership of benefits, a balance of benefits for every benefit owner and the changes required to realize them (Ward & Daniel, 2012). Once the project is in progress benefits may change over time as new issues and working practices emerge. This is because the organisational context in which the new project takes place is likely to have a significant impact on the attitudes and actions of the stakeholders. For this reason business benefits resulting from new IS/IT projects continually evolve and the organization and its context is critical in determining an emerging set of potential benefits. Optimal benefits are those benefits that are aligned with the investment objectives, address strategic stakeholder needs and reference to existing performance management evaluation standards.

To date, there is limited empirical evidence investigating the methods and processes used in practice that define IS/IT benefits (See for e.g. Bennington & Baccarini, 2004). In particular there is a lack of empirical research on the IS/IT benefits management domain that explores the practice of identifying the contextually determined, optimum level of benefits that serve to constitute an organisation's investment objectives. And, equally, even less research investigating these practices in other sub domains such as IT service management (Mcloughlin, Scheepers, & Wijesinghe, 2014). To address this lack this paper directs attention to the following research question:

*What are the processes by which organisations initially define IS/IT benefits?*

Benefits identification is a critical step in the benefit management process, which seeks to identify and document the benefits that will be the most relevant and convincing to contextually impacted stakeholders. One potentially rewarding way to investigate the benefit identification process is by means of the practice based approach. The decomposition of the benefit identification process into a number of constituent practices will allow us to add granularity to the benefit management identification process.

This paper provides insight into a descriptive case study of IS practices adopted by large Fortune 500 organisation in constructing a benefit realization management plan. A single case study strategy will illuminate specific identification practices adopted and help to uncover aspects such as: Why a particular practice was chosen by the case organization?; How the specific practice delivers- unique benefits? Moreover, by focusing on service management project, this research will also try to uncover valuable insights into specific practices of contemporary IS/IT projects. Service management projects are initiated when an organisation needs to improve its delivery of IT services to the business and its customers, and involve both the selection of technology tools as well as the design or re-engineering of existing service management processes that will enable the delivery of quality IT services to the business. (Jäntti, Rout, Wen, Heikkinen, & Cater-Steel, 2013). In this regard, the use of a structured benefits identification framework can improve the potential for the successful implementation of service management tools and improved processes. Therefore the purpose of this paper is to add to the limited body of empirical data on the benefits identification practices adopted by the large case organization and the role the specific investment context plays in shaping these practices.

The paper is structured as follows. We review related work on IS/IT benefits management by focusing on the ITSM domain. We then describe the research methodology and case analysis. Finally we present the discussion arising from the case analysis and summarise this in conclusion.

## 2. RELATED WORK

Our study uses a 'practices' lens which is frequently used in the organisational literature (Wenger, McDermott, & Snyder, 2002), in order to decompose the benefit identification process. Even though studies in Information Systems research have attempted to use a practice based approach to explore capabilities associated in benefit planning (Ashurst, Doherty, & Peppard, 2008), there is little empirical research in the ITSM domain using these academic prescriptions to explore practices relevant to the service management context. According to Wenger et al. (2002) practice is a set of socially defined ways of doing things in a specific domain. In other words, they are common approaches and shared standards that create a basis for action, problem solving, performance and accountability (Wenger et al., 2002). On the other hand, practices are underpinned by the skills, knowledge and experiences of organizational participants and are more concrete and observable (Krogh, Haefliger, Spaeth, & Wallin, 2012; Newell, Tansley, & Huang, 2004). The notion of practice is therefore useful to explore how work is actually done by individuals and groups in a specific context to identify project related benefits (Ashurst et al., 2008).

Although the literature in the practice based approach to benefit planning in the IS/IT context is very limited, three pieces of research are regarded as seminal in the identification of relevant practices in the ITSM context. Firstly, the process model introduced by Ward and Daniel (2012) in benefit management domain provides the necessary structure to decompose benefit realization management related activities. Secondly, the practice based approach adopted by Ashurst et al. (2008) for their benefit realization capability model provides primary content and practices relevant to benefit planning. Finally, best practices in the ITSM domain such as ITIL (Taylor, Cannon, & Wheeldon, 2007) provides the necessary context to analyse our case organisation's practices relevant to benefit identification.

## 2.1 IS/IT Benefits

Post the establishment of clear investment objectives, IS/IT projects usually need to identify the expected benefits (not only the immediate project results which satisfies cost, time, and scope constraints) projected to arise if those investment objectives are met. It emphasizes organizational change which is necessary for the realization of benefits from investments. Benefits can be defined as the advantages provided to specific groups or individuals that result from business changes designed to achieve overall investment objectives (Ward, Daniel, & Peppard, 2008; Ward & Daniel, 2012). Business benefits from IT enabled change usually occur as a result of ceasing certain activities, improving certain activities, or undertaking new activities (Ward et al., 2008; Ward & Daniel, 2012). Based upon the literature review, three distinct activities in IS/IT domain were identified as important practices relevant in benefit identification in ITSM context.

**Practice 1: Stakeholder Engagement**

Engagement of key stakeholders in the benefit identification process plays a key role in project success. According to Bennington and Baccarini (2004) identification of benefits is a combined approach of interviews and workshops which involve key stakeholders who can effectively determine what is required from IS/IT investment and what is affordable and possible in a specific organisational context. Ashurst et al. (2008) stressed that conducting a structured, bottom-up analysis of the stakeholders' requirement is one of the key practices that organisations should master in the benefit planning domain. This maximises the likelihood of stakeholder commitment and also demonstrates the importance of the investment to the overall organisation (Ward et al., 2008). But this alone would not guarantee that the IT organisation would identify the optimal level of potential benefits.

Identification of the most effective set of benefits for the purposes of benefit planning remains challenging due to the complexities inherent in IS/IT projects. Firstly, IS/IT projects by nature involve multiple stakeholders with conflicting objectives and these stakeholders rarely agree on common goals (Bennington & Baccarini, 2004). Secondly, IS/IT benefits continuously evolve throughout the lifespan of the project and hence it is difficult to predict a unique set of benefits in advance for planning purposes (Ward, Taylor, & Bond, 1996). Third, most IS/IT benefits tend to be intangible in nature and very difficult to quantify (Ward et al., 2008; Ward & Daniel, 2012).

**Stakeholders in ITSM context**

Investment projects in the ITSM domain consist of a number of stakeholder groups with a large number of roles ranging from executives to help desk operators (Lempinen & Rajala, 2014). Some of the stakeholder roles are limited in scope and specifically relate to one specific ITSM process, whereas others have responsibilities in several different processes. Moreover, one specific stakeholder group may be required to perform different roles at different times or several roles at the same time, depending on the specific situation and the specific process(es) they are interacting with. Therefore these stakeholder groups typically may have multiple and often conflicting objectives and priorities at any point in time (Lempinen & Rajala, 2014). They rarely agree on a set of common aims. For instance, a role called ITSM incident manager often coordinates activities between multiple support groups to ensure adherence to an extant Service Level Agreement (SLA) and drive towards a resolution either through a temporary workaround or a permanent solution if available. On the other hand, another role called ITSM problem manager ensures the efficient flow of problem tickets through the Problem Management process. A group of stakeholders interested in the source of a problem may not be interested in a temporary fix but would rather want a permanent solution. Similar conflicts also can arise between other roles such as ITSM process owners and services owners. If not properly acknowledged at the benefit planning stage these kinds of potential conflicts between stakeholders can cause disruption to the whole project and could potentially lead to the failure of the project. However these tensions can be avoided by involving the help of all stakeholders in the project at an early stage of the benefit identification process (Ward & Daniel, 2012). Active stakeholder involvement allows IS/IT projects to uncover hidden beliefs and ideas of each stakeholder group which in turn may allow organisations to identify, fine tune and focus common interests. This may also help to avoid vague statements of benefits which in turn may lead to the allocation of uncertain responsibilities and accountabilities in managing and delivery of benefits (Lin & Pervan, 2003).

### Practice 2: Use of Best Practices Guidelines

ITSM is considered a strategy which focuses on defining, managing, and delivering of IT services (Winniford, Conger, & Erickson-Harris, 2009). It helps IT organisations to focus on customer requirements and make the IT department more cost effective, flexible, adaptive, and service oriented. To guide IT organisations to implement ITSM strategy more effectively, the ITIL framework has evolved as the most commonly used best practice guide (Marrone, Gacenga, Cater-Steel, & Kolbe, 2014). These best practices guideline frameworks offer fruitful information on service management processes and organisational structures that have proven effective, rather than guidance on how to apply them (See for e.g. Taylor et al., 2007). These guidelines provide the IT department with an opportunity to customize their ITIL framework to develop its own repeatable, standardized, and documented ITSM processes (Jäntti et al., 2013). This customizable feature allows IT organisations to implement ITSM processes in a way that fits with the requirements and objectives of the organisation. According to Lyytinen and King (2006) IT standards such as ITIL are increasingly important in developing and managing IT services as IT have become ubiquitous, heterogeneous, networked, and complex. Hence, referring to industry wide standards such as ITIL provides an opportunity to identify the unique benefits resulting from the changes the organisation is making with reference to these best practices.

### Practice 3: Use of Benchmarking Partners

Benchmarking is considered one of the ways to measure and compare one organisation against another in order to identify and implement improvements (Andersen & Pettersen, 1995). In the context of benefit realization management, benchmarking can be used to compare internally and externally to learn lessons from similar investment projects and to identify an optimum set of benefits. Internal benchmarking focuses on ITSM processes and takes place between departments or locations of the organisation (Andersen & Pettersen, 1995). External benchmarking focuses on comparing ITSM processes with other organisations (Andersen & Pettersen, 1995). Benchmarking activities highlights problem areas that the organisation needs to pay more attention to and potentially improvement. The benchmarking provides an incentive to all stakeholders to engage with the necessary changes, and assist them to identify benefits and setting targets.

### 2.2 'Disbenefits' in IS/IT Projects

Ward and Daniel (2012) pointed out that potential 'disbenefits' of an IS/IT investment should also be considered. 'Disbenefits' are those adverse impacts on the business and/or organisation and they are the potential 'price worth paying to obtain positive benefits' (Ward & Daniel, 2012). Some outcomes of ITIL implementation may be favourable for the organisation as a whole but perhaps unfavourable for other parts. Any such 'disbenefits' can be identified and tracked by employing a variety of techniques such as involvement of stakeholders, referencing best practices guidelines and benchmarking so their impact can be anticipated and minimised.

By adhering to best practices recommendations, organisations can plan, design and manage IT services in a way that it delivers a variety of ITSM benefits, some of which, may not be identified by stakeholders in advance. But customizability also poses a risk of identifying inappropriate sets of benefits which may not represent the proposed organisational change. This requires organisations to actively engage stakeholders and use benchmarking activities to re-shape identified benefits to reflect the organisation's own investment context and change conditions.

## 3. RESEARCH METHODOLOGY

The findings in this paper are premised on a single case study conducted by researchers at a financial services organisation from July 2013 to December 2014. This case study research was part of a larger longitudinal research project of ITIL implementation in a financial services organisation. The larger project was conducted by an interdisciplinary research team. The larger research project was designed to trace the organisational change process that resulted from an ITIL simplification and implementation project as it unfolded over time. This specific case study research was designed to take advantage of a particularly interesting process within this greater organisational change process. The researchers sought to understand the dynamics of a variety of practices underpinning a benefit identification process that specifically sought to address the IS/IT investment objectives of the project. In this respect it can be regarded as a unique case in that it represented a rather unique approach to the practice of benefits identification. The case study method was used to seek an in-depth understanding of this temporal and multi-faceted organisational process of identifying the IS/IT benefits used to achieve the organisation's investment objectives (Yin, 1994). The case study relies on qualitative data. The case study methodology is a preferred research methodology when "how" questions are asked (Yin, 1994). Close examination of a single case study enables us to understand the mechanics of how an organisation is affected by a variety of factors (Lee, 1989). The in-depth case study method is thought to be appropriate in view of the relatively understudied nature of the research area.

Data was collected from the project initiation phase by means of semi-structured interviews (guided by an interview protocol) and observation with documents as supplementary sources. Data was collected over the period July 2013 to December 2014. Over this period interviews occurred with all members of the project implementation team. Each of the interviews lasted from one to one and a half hours. All interviews were transcribed for analysis. The interviews sought contextual information whilst observing the development of benefit identification processes in the natural context of work. Further, observation occurred at fortnightly steering committee meetings and presentations over the period. Notes were written down either during or shortly after the meetings.

Documentation concerning the IT service simplification plan, the communication plan, the organisational stakeholder analysis, documentation reporting on the identification of benefits and alignment with stakeholder groups, documentation reporting on external organisations experiences with ITSM implementation and benefit identification, stakeholder meeting reports and all meeting notes were collated for thematic analysis in order to document the benefit identification process and outcomes. These documents were identified either on the basis of their relevance to the benefit identification process or from the greater project environment that dealt with overall investment objectives or stakeholder management. Initial coding was undertaken by one researcher and discussed on a fortnightly basis with the research team.

An explication of the case study starts with an introduction to the case organisation and its background, followed by a description of the different benefit identification practices in its benefit planning process.

### 3.1 Overview of the Research Case Organisation

The case company, C Company, is a large Fortune 500 organisation which owns 24 strategic business units (SBUs) around the world. C Company provides a variety of commercial lending and leasing services for large businesses operating in health care, media, communications, entertainment, real estate, and aviation. C Company also provides consumer finances through its retail finance arm, B bank.

C Company used nearly 200 ITSM processes globally to take strategic advantage in each geographical market segments. But after the global recession in 2008, the Federal Reserve required C Company to adopt a more controlled approach to its ITSM function which required a consolidation of the diversified ITSM processes into standardized processes to minimize the overall IT risk exposure. Moreover, due to rapid market change, C Company was compelled to adopt a centralized ITSM function to give top management a single view into service operations. The existing diversified processes did not provide a single customer experience.

### 3.2 Organisational structure of case company

C company maintained a matrix organisational structure in order to enable local market responsiveness and global product/ service consistency. Therefore many IS managers play dual roles at C Company. Parallel reporting relationships at Company C reflected the diverse and conflicting needs of different functional, product and geographical organisational groups.

There are several advantages to C Company's matrix organisational structure. Firstly, this design enhances communication and commonality of purpose among IS managers. For instance, C Company's global CIO (Chief Information Officers) directly reported to the CEO and corporate CIO which manages global Infrastructure Shared Services and Corporate Information Systems divisions. Secondly, the matrix structure allowed C Company to flexibly use its human resources. For instance various individual managers in different IS functional departments were assigned to and held accountable for specific projects such as ITSM Simplification. Thirdly, this matrix structure offered C Company efficient means of responding quickly to the changing and unstable environments post the global financial crisis.

## 3.3 ITSM Process Standardization and Change Initiative (IPSCI) Project

The IPSCI project structure reflects the usual top down hierarchy. The Chief Technology Officer (CTO) of the Australian and New Zealand subsidiary headed up the process excellence team. There were four direct reporting positions under the Process Excellence Leader, namely, ITIL Program Leader for configuration and reporting, ITIL Program Leader for Incident and Portal, ITIL Program Leader for Service Request and User Experience and, Project Leader leading the Change Management strategy team.

The IPSCI project started with a maturity assessment on its ITSM processes in all 24 SBU's by a third party consultancy firm. The assessment found that company's ITSM processes maturity levels were varied among SBUs other than the ITSM Change Management process which was well defined in most of the SBUs. ITSM operational processes such as Incident Management, Problem Management, Request Fulfilment, Knowledge Management, availability management and configuration management were either ad-hoc or not defined in several SBUs. The assessment documentation reveals the following important information on existing ITSM processes and technology:

- The majority of the Incident Management processes in SBUs used a software tool call ServiceNow.
- Knowledge Management process is considered as integral part of Incident Management process at some of the SBUs.
- There were multiple service desks with multiple entry points for incidents.
- Formal verification and audits were not performed in several process areas, specifically in Change Management, Incident Management and Problem Management.
- Some configuration items in configuration management database (CMDB)[1] were outdated.
- Some SBUs did not meet service level agreements (SLAs) in Request Fulfilment and Problem Management. Hence there were long backlogs.
- Limited knowledge sharing between ITSM processes globally.
- Several SBUs use multiple core ITSM tools which were not integrated with each other. For this reason there were more than 3 instances of the ServiceNow system.
- Business units use different ServiceNow forms, fields and maintain their own process documentation.

## 3.4 Practices Adopted by Case Company

As part of the IPSCI project, a number of initiatives were undertaken to generate ideas and to develop strategies for change. Improvements in benefit realization usually comes from putting ideas and knowledge of multi-disciplinary project teams into practices and approaches that enable them to work together more effectively (Ashurst et al., 2008). Therefore in the next section we describe a variety of benefit identification practices adopted by our case company.

### Use of best practices guidelines, publications and benchmarking to identify benefits

As a first step, the IPSCI project team referenced ITIL service operation and service transition publications, and accessed itSMF forum discussions and related white papers to identify an initial list of potential benefits[2]. Secondly, C Company surveyed five comparable organisations and one internal

---

[1] CMDB maintain information about IT assets and the relationships.
[2] itSMF is a non-profit organization dedicated to promote best practices in ITSM. .

SBU who had a high maturity level of ITSM processes as a first step benefit identification and benchmarking exercise. This was to capture types of process related benefits and measures that other comparable organisations have identified in similar ITIL implementation programs.

**Involvement of ITSM stakeholders to elaborate benefits**

As part of stakeholder analysis, the project team identified seven stakeholder groups across the SBUs as primary candidates for benefit identification exercise. They are: CIOs in different SBUs, ITSM Leaders, Chief Technology Officers, ITIL Process Owners, ITIL Process/ Service Managers, Technical Team Members and general Users of IT services. These stakeholders are considered "benefit owners" in the sense that their understanding and buy-in to the project objectives would ensure that the relevant business and enabling changes progress according to plan.

The project team organized a series of workshops and conference meetings involving members from each of the stakeholder groups. The main idea behind these meetings was to encourage open communication among members to identify potential benefits and issues that they would like to address through the ITIL exercise. These workshops and conference meetings were also used to refine the various benefits identified through the methods described above and improve operations and solve existing problems. These meetings and workshops were conducted in an iterative fashion so doubts and problems of various ITSM stakeholders were fined tuned by revisiting ITIL publications, itSMF forum discussions and other publications. After 3 iterative workshops and meetings the project team has constructed a more realistic basket of project benefits. .

## 4. DISCUSSION

The BRM method advises starting with the drivers and strategic objectives of the IS/IT investment and then define what drivers the organisation needs for change. From this the desirable outcomes of this change – the benefits – would then be identified. Case studies reported in the literature describing IS/IT BRM uses the idea of 'practices' to discuss what individuals and groups in IS/IT project setting actually do to enhance the strategic potential of IT (Ashurst et al., 2008). These case studies provide a fresh perspective on the challenges that benefits realization programmes faces and explores these challenges from socio-technical and benefit driven perspectives. The case study described in this paper uses the same approach but we contextualize our investigation into the ITSM context to explore these complexities and how the case organisation adopt different practices to identify most effective list of initial benefits for benefit realization program purpose.

The case organisation was highly dependent on stakeholder participation as part of the benefit identification process through emphasizing cross-functional thinking. This is an important practice adopted by C Company because the final basket of benefits which it identified shows that the interdisciplinary project team went beyond just ensuring that all ITSM stakeholders in the many process areas have a voice. It also ensured that the project team promote ITIL best practices in conjunction with the business needs of vastly complex environments. The result is a collaborative and adaptive list of benefits contributing to a successful benefit realization plan acceptable to all ITIL process users. The objective of stakeholder participation is to obtain buy-in and facilitate change by identifying the obvious benefits and disbenefits resulting from the change, while not threatening local autonomy. On the other hand, moving to a service management framework such as ITIL can be viewed as an organisational cultural change. For example, existing service management processes appear to support the existing decentralized structure, but the adoption of an ITIL framework appears to strengthen the centralized authority structure. In this respect developing a clear understanding of the potential benefits for all relevant stakeholders and the organisation as a whole is a very important practice for a successful and smoother transition.

Identification of benefits for the purpose of benefit planning is, or should be, an ongoing activity through the project lifespan (Ward & Daniel, 2012). In situations where there is a greater element of innovation, there will be a need to revisit the basket of benefits which was identified in the first iteration. The understanding of the problem and/or opportunity will have to evolve and there will be fresh insights into potential benefits from key stakeholder groups. Our case organisation went through three iterative cycles to understand the initial list of benefits. At the initial iteration, the project team accepted the importance of the IPSCI project to achieve maximum benefits from the new ITIL adoption and the importance of having a high degree of certainty for the benefit plan. The project team then established key stakeholder groups to include all relevant knowledge and information and gave the opportunity to stakeholders to understand the implications of the benefit plan and their roles in its execution. In order to facilitate knowledge sharing between project sponsors and the ITSM/ITIL specialists, the project team also established a series of workshops rather than holding meetings or one-to-one discussions. Overall these workshops have ensured that the IPSCI project links the necessary business drivers with the overall benefits and which can be sustained in the longer run. It also ensured that the relationship between benefits and changes are made explicit. Between iterations, a considerable amount of work was done by both project team members and other key stakeholders. These activities included documenting best practices benefits and refining it with current project circumstances and reviewing benchmarking interviews carried out by the project team to obtain further information. It also included the reviewing of outputs from the first iteration.

Before the start of the second iteration, the project team ensured that all stakeholder implications of the IPSCI project had been identified as were their perceptions about how it will affect them and their consequent ability and willingness to fulfil their expected responsibilities. This also included the stakeholders view of any 'disbenefits' that may result from the investment, so that any concerns and potential resistance would be identified and considered at the next high level workshop. Additionally, the project team defined measurements and quantify the proposed benefits.

During the second and third iterations, the project team ensured that they estimated the expected benefits and the feasibility of providing the required ITSM/ITIL components (mainly IT tools). The team realised that some parts of the project provide significant benefits whilst other changes were deemed too risky. For instance, the proposed ITSM intelligence tool was deemed to be too risky to implement because there was not sufficient evidence to prove that the new solution will be acceptable to all stakeholder groups. In any project it is important to maintain the momentum and keep the key stakeholders involved and interested in the benefit identification (Ward & Daniel, 2012). Although those iterations brought stakeholders with different knowledge sets together to agree what needs to be done, it also acted as a stepping stone to identify useful 'tips' or guidance on how to make them work together successfully.

The case study clearly illustrate that the use of various processes has contributed towards creating a common understanding of potential benefits the IPSCI project could generate. The simplification of the ITIL process meant that disparate stakeholders came together and agreed on common benefits. The main processes used to develop this common understanding and their contributions are illustrated Table 1 below:

| Processes | Contribution | Result |
|---|---|---|
| Best practise guidelines from ITIL publications | The literature on ITIL implementation developed a general understanding of what potential benefits a simplified ITIL implementation could provide | A list of industry benefits was identified. |
| Best practise guidelines from comparable organisations | The survey of comparable organisations and one internal SBU provided process related benefits and measures for ITIL implementation projects | A refined list of benefits was generated. A list of measures was identified. |
| Benefit identification by 'benefit owners' | A series of workshops and conference meetings were held to identify issues and benefits | A final list of benefits was identified through the various workshops through active stakeholder involvement. |

*Table 1. Main processes used to identify benefits*

## 5. CONCLUSION

Even though BRM approach is fast gaining traction within IT organisations around the world, there appears to be limited attention paid to the intra-organisational practice by which such benefits are identified. Our paper describes a case organisation's adoption of BRM programme which uses a number of benefit identification processes in an effort to more comprehensively identify IS/IT related benefits that reflect the organisational investment. We also discuss the importance of this benefit identification process in the context of ITSM. This is achieved through the case examination of an information technology infrastructure library (ITIL) implementation in a multi-national organisation. The case analysis shows that use of practices such as best practices guidelines, benchmarking partners and active stakeholder involvement has been an important success factors in the achievement of a realistic benefit plan.

The research described above is the first phase in a larger research project into ITIL implementation in large organisations. The case study describes a financial organisation that has to change due to external pressure. Further research is necessary to determine if other large financial or non-financial organisations not under the same external pressure would embark on the same processes and would find similar results in the use of these processes. In addition the role of organisational culture needs to be further explored as this could play a significant role in the willingness of stakeholders to fully engage in the processes. The next step in the project is the investigation of the ITIL implementation processes in additional organisations.

## 6. REFERENCES

Andersen, B., & Pettersen, P. G. 1995. *Benchmarking Handbook*: Springer.

Ashurst, C., Doherty, N. F., & Peppard, J. 2008. Improving the impact of IT development projects: the benefits realization capability model. *European Journal of Information Systems*, 17(4): 352-370.

Bennington, P., & Baccarini, D. 2004. PROJECT BENEFITS MANAGEMENT IN IT PROJECTS-AN AUSTRALIAN PERSPECTIVE. *Project Management Journal*, 35(2): 20-30.

Gacenga, F., Cater-Steel, A., Toleman, M., & Tan, W.-G. 2011. Measuring the performance of service orientated IT management. *Sprouts: working papers on information environments, systems and organizations*, 11(162).

Gingnell, L., Franke, U., Lagerström, R., Ericsson, E., & Lilliesköld, J. 2014. Quantifying Success Factors for IT Projects—An Expert-Based Bayesian Model. *Information Systems Management*, 31(1): 21-36.

Jäntti, M., Rout, T., Wen, L., Heikkinen, S., & Cater-Steel, A. 2013. Exploring the Impact of IT Service Management Process Improvement Initiatives: A Case Study Approach. In T. Woronowicz, T. Rout, R. O'Connor, & A. Dorling (Eds.), *Software Process Improvement and Capability Determination*, Vol. 349: 176-187: Springer Berlin Heidelberg.

Krogh, G. V., Haefliger, S., Spaeth, S., & Wallin, M. W. 2012. Carrots and rainbows: motivation and social practice in open source software development. *MIS Q.*, 36(2): 649-676.

Lee, A. S. 1989. A Scientific Methodology for MIS Case Studies. *MIS Quarterly*, 13(1): 33-50.

Lempinen, H., & Rajala, R. 2014. Exploring multi-actor value creation in IT service processes. *Journal of Information Technology*, 29(2): 170-185.

Lin, C., & Pervan, G. 2003. The practice of IS/IT benefits management in large Australian organizations. *Information & Management*, 41(1): 13-24.

Lyytinen, K., & King, J. L. 2006. STANDARD MAKING: A CRITICAL RESEARCH FRONTIER FOR INFORMATION SYSTEMS RESEARCH. *MIS Quarterly*, 30: 405-411.

Marrone, M., Gacenga, F., Cater-Steel, A., & Kolbe, L. 2014. IT service management: a cross-national study of ITIL adoption. *Communications of the Association for Information Systems*, 34(1): 865-892.

Mcloughlin, S., Scheepers, H., & Wijesinghe, R. 2014. Benefit Planning Management for ITSM: Evaluating Benefit Realization Frameworks, *The 25th Australasian Conference on Information Systems*. Auckland ACIS.

Newell, S., Tansley, C., & Huang, J. 2004. Social Capital and Knowledge Integration in an ERP Project Team: The Importance of Bridging AND Bonding. *British Journal of Management*, 15: S43-S57.

Taylor, S., Cannon, D., & Wheeldon, D. 2007. *Information Technology Infrastructure Library (ITIL) version 3 (refresh) - Service operation*. London: The Stationery Office.: Office of Government Commerce (OGC).

Ward, J., Daniel, E., & Peppard, J. 2008. BUILDING BETTER BUSINESS CASES FOR IT INVESTMENTS. *MIS Quarterly Executive*, 7(1): 1-15.

Ward, J., Taylor, P., & Bond, P. 1996. Evaluation and realization of IS/IT benefits: an empirical study of current practice. *European Journal of Information Systems*, 4(4): 214-225.

Ward, J. L., & Daniel, E. 2012. *Benefits management* (2nd ed.). Chichester: Wiley.

Wenger, E., McDermott, R. A., & Snyder, W. 2002. *Cultivating Communities of Practice: A Guide to Managing Knowledge*: Harvard Business School Press.

Winniford, M., Conger, S., & Erickson-Harris, L. 2009. Confusion in the Ranks: IT Service Management Practice and Terminology. *Information Systems Management*, 26(2): 153-163.

Yin, R. K. 1994. *Case study research: design and methods*: Sage Publications.